\documentclass{article}

\usepackage[final, nonatbib]{neurips_2024}
\usepackage{amsmath}
\usepackage{graphicx}
\usepackage{float}

\usepackage[utf8]{inputenc} 
\usepackage[T1]{fontenc}    
\usepackage{hyperref}       
\usepackage{url}            
\usepackage{booktabs}       
\usepackage{amsfonts}       
\usepackage{nicefrac}       
\usepackage{microtype}      
\usepackage{xcolor}         

\title{Solaris: A Foundation Model of the Sun}

\author{Harris Abdul Majid$^*$ \\
    The University of Edinburgh \\
    \texttt{h.abdulmajid@ed.ac.uk} \\
    \And
    Pietro Sittoni \thanks{Equal contribution} \\
    Gran Sasso Science Institute \\
    \texttt{pietro.sittoni@gssi.it} \\
    \And 
    Francesco Tudisco \\
    The University of Edinburgh \\
    \texttt{f.tudisco@ed.ac.uk} \\
}

\begin{document}
\maketitle

\begin{abstract}
Foundation models have demonstrated remarkable success across various scientific domains, motivating our exploration of their potential in solar physics. In this paper, we present Solaris, the first foundation model for forecasting the Sun's atmosphere. We leverage 13 years of full-disk, multi-wavelength solar imagery from the Solar Dynamics Observatory, spanning a complete solar cycle, to pre-train Solaris for 12-hour interval forecasting. Solaris is built on a large-scale 3D Swin Transformer architecture with 109 million parameters.
We demonstrate Solaris' ability to generalize by fine-tuning on a low-data regime using a single wavelength (1700 Å), that was not included in pre-training, outperforming models trained from scratch on this specific wavelength. Our results indicate that Solaris can effectively capture the complex dynamics of the solar atmosphere and transform solar forecasting. 
\end{abstract}

\section{Introduction}
Foundation models have achieved state-of-the-art performance across various domains, including natural language processing, computer vision, and robotics \cite{bommasani2021opportunities}. This success has sparked interest among researchers in diverse scientific fields. Recent examples include ESM-2 for biology \cite{lin2022language}, ChemBERTa-2 for chemistry \cite{ahmad2022chemberta}, and Poseidon for differential equations \cite{herde2024poseidon}. In atmospheric science, foundation models like GraphCast \cite{lam2022graphcast}, ClimaX \cite{nguyen2023climax}, and Aurora \cite{bodnar2024aurora} have garnered significant attention, demonstrating the ability to match or even exceed the accuracy of state-of-the-art numerical weather prediction methods while requiring only a fraction of the compute. 

Building on the success of foundation models in atmospheric science, we propose training a foundation model on the Sun's atmosphere. The Sun's atmosphere is a vast plasma environment extending millions of kilometers into space, comprising several layers including the photosphere, chromosphere, and corona. Each layer is characterized by an increase in temperature and a decrease in density as the distance from the Sun's surface increases. The Sun's atmosphere is crucial in solar-terrestrial interactions, significantly impacting Earth and Earth's atmosphere. Solar flares and solar winds can trigger geomagnetic storms on Earth, disrupting power grids, satellite communications, and space missions. Additionally, solar irradiance influences climate modeling and long-term weather forecasting. Given these far-reaching effects, developing accurate models to forecast the solar atmosphere would contribute to protecting critical infrastructure and advancing scientific understanding.

To date, deep learning applications for the solar atmosphere have focused on specialized models for specific tasks, such as convolutional neural networks for solar flare classification \cite{zheng2019solar, li2020predicting}, or recurrent neural networks for solar parameter forecasting \cite{muralikrishna2022exploring}. However, the use of foundation models for solar atmospheric research remains largely unexplored. These models could be pre-trained on diverse data types, including spectral and photometer measurements, dopplergrams, magnetograms, and full-sun images across multiple wavelengths, enabling them to learn generalizable representations of solar atmospheric processes.

Foundation models for the Sun's atmosphere could offer significant advantages over their task-specific counterparts. By incorporating a wide range of solar atmospheric phenomena, these models could develop a more comprehensive understanding of fundamental processes. Furthermore, they could be fine-tuned for various downstream tasks, potentially achieving enhanced performance. This approach aligns with the successful application of foundation models in other scientific domains, including weather and atmospheric prediction.

In this paper, we present Solaris, a foundation model of the Sun. The remainder of this paper is structured as follows: In Section~\ref{sec:data}, we provide details of the data. Then, in Section~\ref{sec:model}, we discuss the model architecture, comprising an encoder, a processor, and a decoder, along with the image normalization and transformation steps. Finally, in Section~\ref{sec:experiments}, we present pre-training and fine-tuning experiments and show that Solaris holds promise for effectively modeling and forecasting the Sun's atmosphere.

\section{Data}\label{sec:data}
The Sun's atmosphere is observed by various instruments, both on Earth and in space, generating vast amounts of data for heliophysics research. A significant source of this data is the Solar Dynamics Observatory (SDO), launched in 2010, which generates several terabytes of observational data daily through its three main instruments:
\begin{enumerate}
    \item The Extreme Ultraviolet Variability Explorer (EVE) measures the Sun's extreme ultraviolet irradiance with high spectral resolution and temporal cadence.
    \item The Helioseismic and Magnetic Imager (HMI) provides full-disk dopplergrams and magnetograms of the Sun's photosphere at 0.6 arcsec/pixel resolution (4096x4096 pixel images) every 45 seconds.
    \item The Atmospheric Imaging Assembly (AIA) captures full-disk images of the Sun's chromosphere and corona at 0.6 arcsec/pixel resolution (4096x4096 pixel images) in 10 ultraviolet and extreme ultraviolet wavelengths every 10 seconds.
\end{enumerate}
The Joint Science Operations Center (JSOC) performs low-level processing and calibration for all HMI and AIA images. After processing, JSOC provides several processed datasets, including synoptic data. For HMI, the synoptic dataset includes downsampled 1024x1024 pixel images (2.4 arcsec/pixel), sampled once per hour. For AIA, the synoptic dataset includes downsampled 1024x1024 pixel images (2.4 arcsec/pixel) of each wavelength, sampled once every two minutes.

For our study, we focused on AIA synoptic data. Specifically, we utilized all available wavelengths from July 1, 2010, to December 31, 2023, sampling at 12-hour intervals (00:00 and 12:00 UTC daily). This dataset spans a complete 11-year solar cycle, crucial for capturing the full range of solar activity, including both quiet and active solar conditions, allowing for improved generalization across various solar activity levels.

We followed the preprocessing steps outlined in \cite{galvez2019machine}. First, we applied degradation correction to account for sensor deterioration over time and exposure correction to standardize all images to a 1-second exposure time. We then centered and scaled each image to a target angular size of 976.0 arcseconds to provide a uniform field of view. Finally, we resized the images to 512x512 pixels.

One of the main challenges in this study involved acquiring and preprocessing the data. For some time-wavelength pairs, data files were missing or unavailable, resulting in an uneven number of samples across wavelengths. Most wavelengths had approximately 9000 samples, while the 1700 Å wavelength had only 987 samples. Additionally, some existing data files contained only metadata without image data. Occasionally, data files containing image data were of poor quality, likely due to regular calibration activities interrupting AIA's standard operation, resulting in images that were dark, off-center, or out of focus. To maintain dataset quality, we inspected the \texttt{QUALITY} keyword in the metadata to identify and discard poor-quality images.

We preprocessed and stored the dataset in \texttt{.hdf5} files for ease of handling in machine learning tasks and made it publicly available. \footnote{\url{https://huggingface.co/datasets/hrrsmjd/AIA_12hour_512x512}. Although our final dataset is over 100GB, it represents only $\sim$ 0.3\% of the available AIA synoptic data, highlighting the vast amount of solar data available and the potential for future work with even larger datasets.} To our knowledge, this represents the first publicly available machine learning dataset for solar atmospheric studies that covers a full solar cycle, offering a unique resource for researchers in heliophysics and machine learning. 

\begin{figure}
    \centering
    \includegraphics[width=1\linewidth]{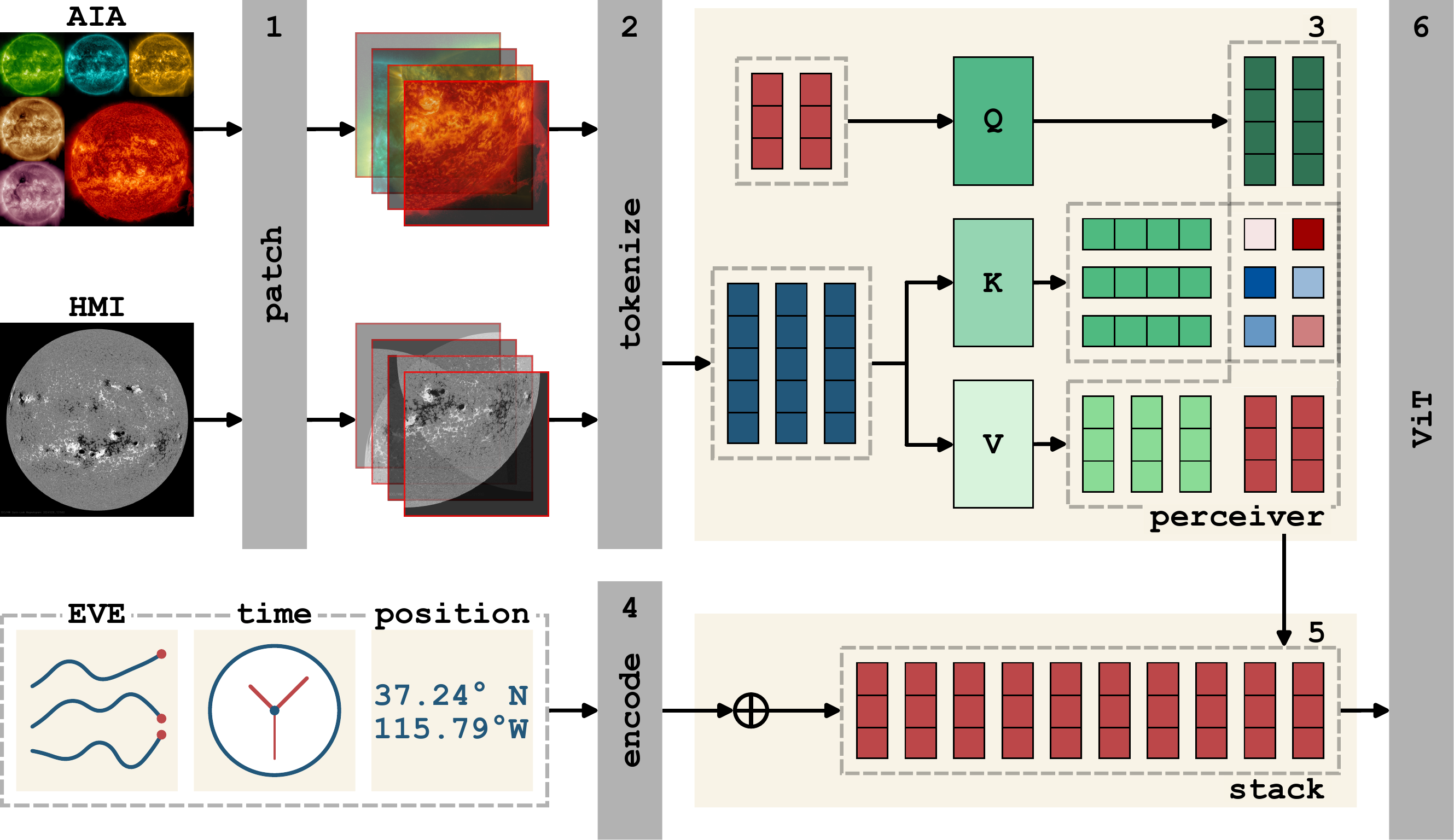}
    \caption{Illustration of Solaris' encoder module. The encoder processes multi-wavelength solar observations from AIA, HMI, and EVE instruments through patch embedding, tokenization, perceiver-based aggregation using cross-attention, spatiotemporal encoding, and transformer stack processing. The encoded representation is then passed to the Swin Transformer (ViT) processor for temporal evolution modeling.}
    \label{fig:model}
\end{figure}

\section{Model}\label{sec:model}
Building on the success of Aurora in atmospheric forecasting \cite{bodnar2024aurora}, we developed Solaris, a foundation model for solar atmospheric forecasting. Solaris closely follows Aurora's architecture, adapting it for the unique challenges of solar physics. Like its terrestrial counterpart, Solaris consists of three main components:
\begin{enumerate}
    \item A perceiver-based \cite{jaegle2021perceiver} \textbf{encoder} that maps full-disk, multi-wavelength solar imagery into a standardized 3D representation of the Sun's observable atmosphere.
    \item A Vision Transformer (ViT) \cite{alexey2020image} \textbf{processor} that evolves this representation through time, capturing the complex dynamics of the Sun.
    \item A perceiver-based \textbf{decoder} that transforms the evolved 3D representation back into forecasts for each specified wavelength.
\end{enumerate}
This architecture enables Solaris to accept and predict solar atmospheric states across multiple wavelengths and varying spatial resolutions, mirroring Aurora's flexibility in handling diverse atmospheric data.

\subsection{Encoder}
Solaris' encoder accepts inputs of shape $B \times T \times C \times H \times W$, where $B$ represents the batch size, $T$ the temporal dimension, $C$ the number of wavelengths, and $H$ and $W$ the spatial dimensions of the images. The encoding process consists of several steps:
\begin{enumerate}
    \item Patch embedding: Each wavelength image is independently divided into $P \times P$ patches. These patches are then embedded into vectors of dimension $\mathbb{R}^D$ using a linear transformation.
    \item Wavelength aggregation: A perceiver-based module utilizes a cross-attention mechanism to condense the variable number of wavelengths into a fixed set of latent representations.
    \item Spatial and temporal encoding: To capture both fine- and coarse-grained patterns in space and time, we apply Fourier encoding: 
    \begin{enumerate}
        \item Positional information: Pixel-level Fourier encoding is added to the latent representation.
        \item Temporal information: We incorporate two types of temporal embeddings:
        \begin{itemize}
            \item Absolute time embeddings: The number of hours since a fixed reference date.
            \item Lead time embeddings: The forecast horizon in hours.
        \end{itemize}
    \end{enumerate}
\end{enumerate}
The output of the encoder is a 3D representation of the solar atmospheric state, capturing both spatial and temporal features across different wavelengths. 

\subsection{Processor}
Solaris makes use of a 3D Swin Transformer U-Net as its processor \cite{liu2021swin, cao2022swin}, an architecture that has proven effective in similar forecasting tasks. The 3D Swin Transformer consists of its own encoder and decoder, each containing multiple stages that progressively adjust the spatial resolution of the 3D representation. Within each layer of the encoder and decoder, local self-attention operations are performed within shifting windows. This design enables efficient information flow across the representation while remaining computationally feasible at high resolutions. The multi-scale structure of the 3D Swin Transformer aligns with the hierarchical nature of physical processes in the Sun's atmosphere and allows Solaris to capture complex, multi-scale dynamics across diverse spatial and temporal scales.

\subsection{Decoder}
Solaris' decoder mirrors the structure of the encoder, applying a series of transformations to reconstruct the original image size. It produces an output of dimensions $B \times C_T \times H \times W$, where $C_T$ is the number of specified wavelengths. The decoder first transforms the latent representations into representations for the specified wavelengths rather than the full set of input wavelengths. These transformed representations are then projected to patch-level predictions using a linear layer before being reassembled into full-resolution images through an inverse patch embedding operation. As a result, Solaris can flexibly output predictions for any specified subset of wavelengths, maintaining consistency with the input structure while effectively translating the abstract latent representations into solar imagery for the desired wavelengths.

\subsection{Normalization and Transformation}
One of the key challenges in working with multi-wavelength solar imagery is the vast difference in intensity scales across different wavelengths. This variation can lead to certain wavelengths dominating the model's training, potentially biasing the model's performance.

To bring all wavelengths to a comparable scale, we computed a scaling factor for each wavelength. This factor is calculated as half the average of the maximum pixel values across all images of that wavelength (in the training dataset). We then normalized each image by dividing all pixel values by this scaling factor. This approach typically maps the pixel values into the range [0, 2] across all wavelengths, creating a more balanced input distribution for the model.

While normalization addresses the scale differences between wavelengths, it doesn't account for the wide range of intensities within each wavelength. Solar imagery often contains regions of high intensity against a background of much lower intensity. To balance the model's sensitivity to both high and low-intensity features, we applied an additional transformation to the normalized data:
\[
\hat{x} = c_1 \min (x, 2.5) + c_2 \frac{\log(10^{-3}) - \log(\max(x, 10^{-3}))}{\log(10^{-3})},
\]
where $x$ is the normalized input, and $c_1$ and $c_2$ are learnable parameters initialized to 0.5. This transformation combines the original normalized values with their logarithm, effectively balancing the network's sensitivity across different magnitudes.

During training and inference, we input these transformed images into the model. When computing the loss, we first unscale the model's output, calculate the loss on the original scale, and then scale the loss. This process ensures that the loss calculation remains meaningful across all wavelengths. By combining normalization and transformation, we enable Solaris to effectively handle multi-wavelength solar data, ensuring balanced learning and sensitivity across the full range of observed intensities.

\subsection{Configurations}
\begin{table}[t]
\caption{Solaris' Configurations.}
\label{solaris-configurations}
\centering
\resizebox{1\columnwidth}{!}{
  \begin{tabular}{lcccc}
    \toprule
    Model      & Model Size   & Backbone Enc Layers & Backbone Dec Layers & Embed Dim \\
    \midrule
    $\text{Solaris}_{\text{T}}$ & 24M & (2, 4, 2) & (2, 4, 2) & 128 \\
    $\text{Solaris}_{\text{S}}$ & 117M & (2, 6, 2) & (2, 6, 2) & 256 \\
    \bottomrule
  \end{tabular}
}
\end{table}

The Solaris model comes in two configurations: Solaris tiny ($\text{Solaris}_{\text{T}}$) and Solaris small ($\text{Solaris}_{\text{S}}$) (Table \ref{solaris-configurations}). Solaris tiny is the more compact version with 24M parameters, using a backbone structure of (2, 4, 2) layers for both encoder and decoder, and an embedding dimension of 128. Solaris small scales up to 117M parameters, with a slightly deeper backbone of (2, 6, 2) layers and an increased embedding dimension of 256. These configurations allow for experimentation with model capacity while maintaining the overall architecture.

\section{Experiments}\label{sec:experiments}
In this section, we evaluate Solaris, focusing on its ability to forecast future states of the Sun's atmosphere based on historical observations. This task is fundamental to understanding and predicting solar dynamics, with significant implications for space weather forecasting and solar physics research.

We formalize the problem as follows: let $X^t \in \mathbb{R}^{C \times H \times W}$ represent the observed state of the Sun's atmosphere at time $t$, where $C$ denotes the number of wavelengths, and $H$ and $W$ are the spatial dimensions of the images. The primary objective of this study is to develop a function F that accurately predicts the future state of the Sun's atmosphere using historical data:
\[
F(X^t, X^{t-1}) = \hat{X}^{t+1} \approx X^{t+1}.
\]
For all experiments in this study, we utilize two historical states with a 12-hour separation to forecast the state 12 hours ahead.

To rigorously assess Solaris' performance and its ability to generalize, we temporally split our dataset as follows: the training set covers data from July 1, 2010, to December 31, 2022; the validation set covers data from January 1, 2023, to June 30, 2023; and the test set covers the data from July 1, 2023, to December 31, 2023. This chronological division ensures that our model is evaluated on future data, closely simulating real-world forecasting conditions.

The subsequent subsections detail our experimental setup and results.

\begin{figure}
    \centering
    \includegraphics[width=1.0\linewidth]{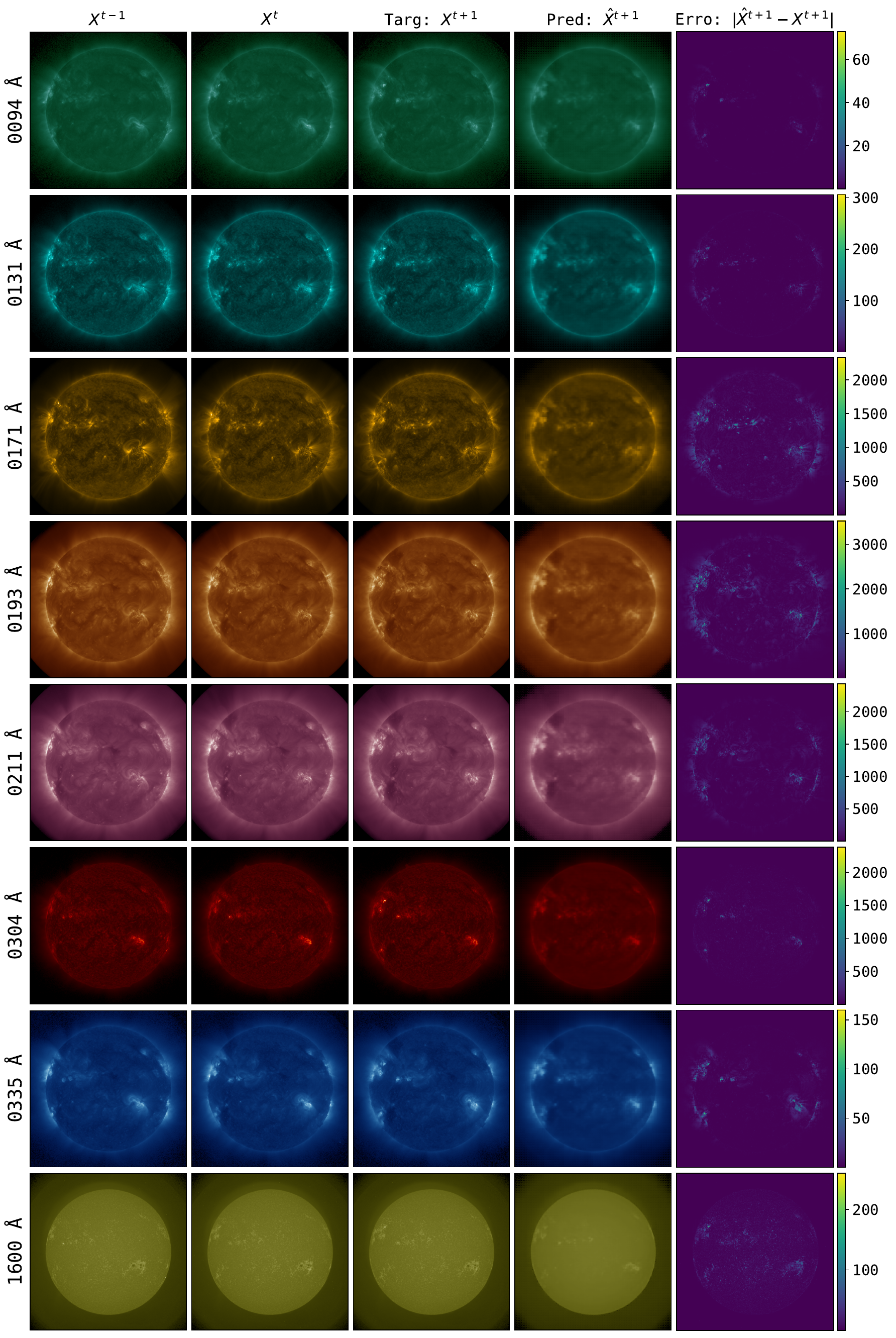}
    \caption{Multi-wavelength forecasting results from $\text{Solaris}_\text{T}$ for 2023-07-02 00:00 UTC. From left to right: input states $X^{t-1}$ and $X^t$, target state $X^{t+1}$, predicted state $\hat{X}^{t+1}$, and absolute error $|\hat{X}^{t+1} - X^{t+1}|$. Rows show results across different wavelengths (94, 131, 171, 193, 211, 304, 335, and 1600 Å).}
    \label{fig:plot_tiny_1}
\end{figure}

\begin{figure}
    \centering
    \includegraphics[width=1.0\linewidth]{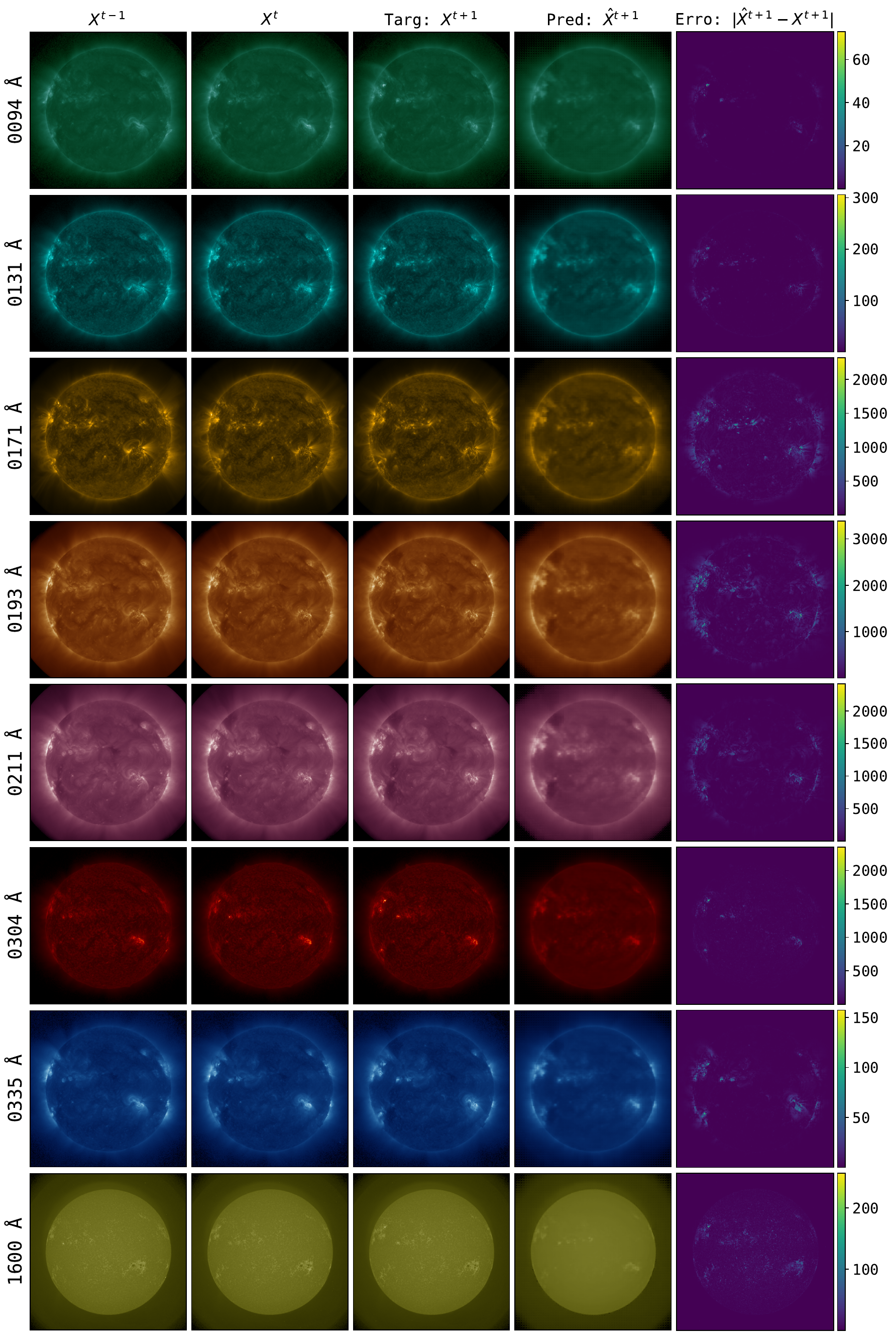}
    \caption{Multi-wavelength forecasting results from $\text{Solaris}_\text{S}$ for 2023-07-02 00:00 UTC. From left to right: input states $X^{t-1}$ and $X^t$, target state $X^{t+1}$, predicted state $\hat{X}^{t+1}$, and absolute error $|\hat{X}^{t+1} - X^{t+1}|$. Rows show results across different wavelengths (94, 131, 171, 193, 211, 304, 335, and 1600 Å).}
    \label{fig:plot_small_1}
\end{figure}

\subsection{Pretraining}
One of the main challenges in this forecasting task lies in the incomplete nature of the observations. The AIA images provide only a partial view of the Sun's atmosphere, capturing data from the SDO-facing side. This limitation is particularly problematic when attempting to predict the emerging regions---areas that are currently on the far side of the Sun but will rotate into view. The feasibility of accurately forecasting these regions is uncertain, as it would require the model to develop an internal representation of the Sun's full atmospheric state, including areas not directly observed. Successfully addressing this challenge would demonstrate Solaris' ability to forecast solar atmospheric dynamics from limited observational data.

\paragraph{Experimental setup.} The model input consists of states at 8 wavelengths (94, 131, 171, 193, 211, 304, 355, 1600 Å) from two time points, separated by 12 hours. The model forecasts states at the same 8 wavelengths for the next time point, 12 hours ahead. We trained the model with a batch size of 8 using 4 gradient accumulation steps, resulting in an effective batch size of 32. Training continued for 7750 steps using the AdamW optimizer. The learning rate schedule featured a linear warmup to 5e-4, followed by a cosine decay to 5e-5, with weight decay set to 5e-6. We used a weighted Mean Absolute Error (MAE) as the loss function.

\paragraph{Results.} Both $\text{Solaris}_{\text{T}}$ and $\text{Solaris}_{\text{S}}$ demonstrate the ability to forecast the Sun's atmosphere across all eight wavelengths (Figures \ref{fig:plot_tiny_1} and \ref{fig:plot_small_1}). Solaris accurately predicts high-intensity regions, not only in currently observable areas but also in emerging sections that rotate into view. This uniform accuracy across observable and emerging regions suggests that Solaris has developed an internal representation of the Sun's global state. Additionally, the model's performance is consistent across all wavelengths, despite the vast differences in scales between them. This indicates that the normalization and transformation techniques effectively balanced Solaris's sensitivity across different wavelengths. The model captures fine details and structures in the solar atmosphere, with predicted states closely matching the ground truth across various features and intensity levels, further demonstrating its forecasting capabilities.

\begin{figure}
    \centering
    \includegraphics[width=0.5\linewidth]{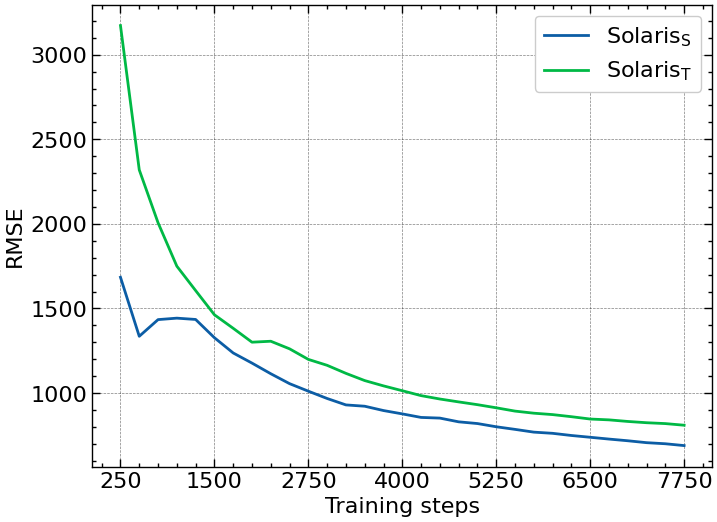}
    \caption{RMSE error during training of $\text{Solaris}_{\text{S}}$ and $\text{Solaris}_{\text{T}}$  }
    \label{fig:rmse_tiny_vs_small}
\end{figure}

The training performance of $\text{Solaris}_{\text{T}}$ and $\text{Solaris}_{\text{S}}$ demonstrates the effectiveness of model scaling in solar atmospheric forecasting (Figure \ref{fig:rmse_tiny_vs_small}). $\text{Solaris}_{\text{S}}$ consistently achieves lower RMSE values throughout the training process. This finding indicates the potential for further enhancements through increased model capacity.

\begin{table}
  \caption{Solaris' Performance (RMSE).}
  \label{rmse}
  \centering
  \begin{tabular}{lcccccccc}
    \toprule
     & \multicolumn{8}{c}{Wavelength (Å)} \\
    \cmidrule(lr){2-9}
    Model & 94 & 131 & 171 & 193 & 211 & 304 & 355 & 1600 \\
    \midrule
    $\text{Solaris}_{\text{T}}$ & $3.023 $  & $15.822$   & $105.862$   & $152.906$   & $75.312$   & $35.805$   & $6.307$   & $8.670$ \\
    $\text{Solaris}_{\text{S}}$ & $\mathbf{2.99}$  & $\mathbf{15.70}$   & $\mathbf{99.579}$   & $\mathbf{138.724}$   & $\mathbf{64.196}$   & $\mathbf{34.295}$   & $\mathbf{5.840}$   & $\mathbf{7.797}$ \\
    \bottomrule
  \end{tabular}
\end{table}

\subsection{Finetuning}
A significant challenge in solar atmospheric forecasting is the uneven availability of data across different wavelengths. Specifically, the 1700 Å wavelength is observed in only 987 of our samples, compared to approximately 9000 observations for other wavelengths. This scenario presents an opportunity to evaluate Solaris' ability to leverage knowledge gained from well-represented wavelengths to improve forecasting for a sparsely observed wavelength. Successfully fine-tuning Solaris for the 1700 Å wavelength would not only demonstrate the model's adaptability but also offer a potential solution for completing missing data in the JSOC archive. This application highlights the practical benefits of a foundation model approach in addressing real-world data limitations in solar physics research.

\paragraph{Experimental setup.} For this fine-tuning task, we maintained the same overall experimental configuration as in the pretraining phase, with a crucial difference in the input and output structure. The input consists of states at the original 8 wavelengths (94, 131, 171, 193, 211, 304, 355, 1600 Å) from two time points, separated by 12 hours. However, the model now forecasts only the 1700 Å wavelength state 12 hours ahead. To evaluate the effectiveness of our foundation model approach, we compared the performance of a fine-tuned $\text{Solaris}_{\text{S}}$ against a model of identical architecture trained from scratch on this task.

\begin{figure}
    \centering
    \includegraphics[width=0.5\linewidth]{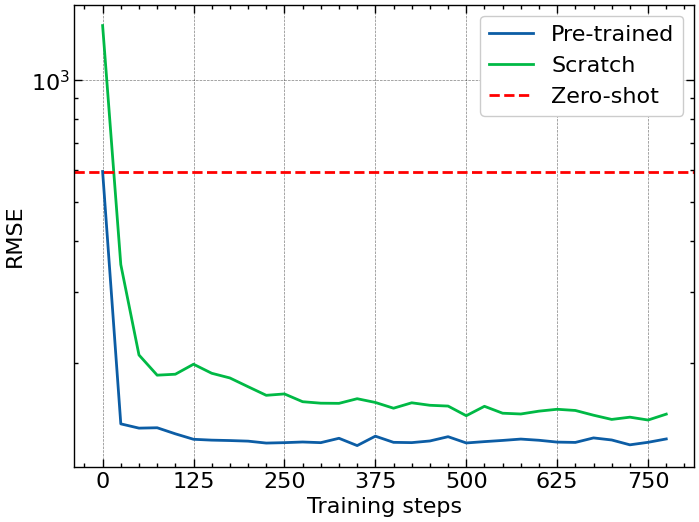}
    \caption{RMSE error during finetuning of $\text{Solaris}_{\text{S}}$.}
    \label{fig:rmse_tiny_vs_small2}
\end{figure}

\paragraph{Results.} Finetuning $\text{Solaris}_{\text{S}}$ for only 25 training steps on the 1700 Å wavelength task, outperformed an identical model trained from scratch for 775 steps. As illustrated in Figure \ref{fig:rmse_tiny_vs_small2}, the pre-trained model's RMSE dropped rapidly and remained consistently lower than the model trained from scratch throughout the fine-tuning process. This result demonstrates the effectiveness of the foundation model approach in solar atmospheric forecasting. The pre-trained model's ability to leverage knowledge gained from other wavelengths enabled superior performance with substantially less task-specific training, highlighting the potential of Solaris to adapt quickly to new, data-limited tasks in solar physics research.

\section{Discussions}
Our results demonstrate Solaris' potential as a foundation model for solar atmospheric forecasting, yet several avenues for improvement remain. Future iterations could incorporate more data, including higher resolution imagery (1024x1024) and additional data types like EVE and HMI observations, potentially enhancing the model's understanding of solar dynamics. Scaling up the model and using finer patch sizes (currently 8x8 due to computational constraints) could improve the capture of fine-scale solar features. Given Solaris' apparent internal representation of the Sun's global state, future research could explore methods to extract and interpret this information, potentially providing insights into solar dynamics on the far side. These directions highlight Solaris' potential to advance solar physics research and forecasting capabilities.

\newpage

\bibliographystyle{plain}


\newpage

\appendix


\begin{figure}
    \centering
    \includegraphics[width=1.0\linewidth]{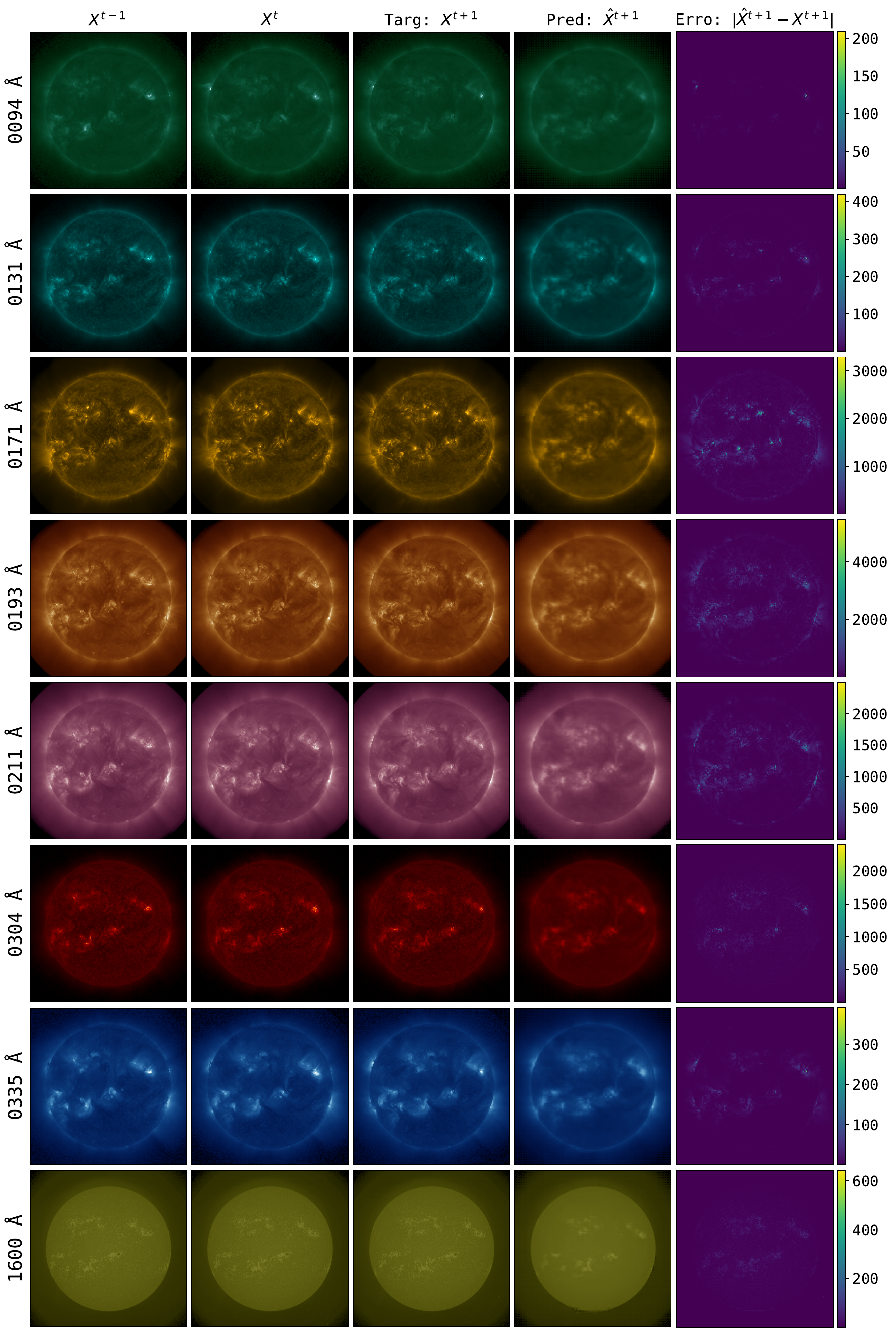}
    \caption{Multi-wavelength forecasting results from $\text{Solaris}_\text{T}$ for 2023-08-02 00:00 UTC. From left to right: input states $X^{t-1}$ and $X^t$, target state $X^{t+1}$, predicted state $\hat{X}^{t+1}$, and absolute error $|\hat{X}^{t+1} - X^{t+1}|$. Rows show results across different wavelengths (94, 131, 171, 193, 211, 304, 335, and 1600 Å).}
    \label{fig:plot_tiny_2}
\end{figure}

\begin{figure}
    \centering
    \includegraphics[width=1.0\linewidth]{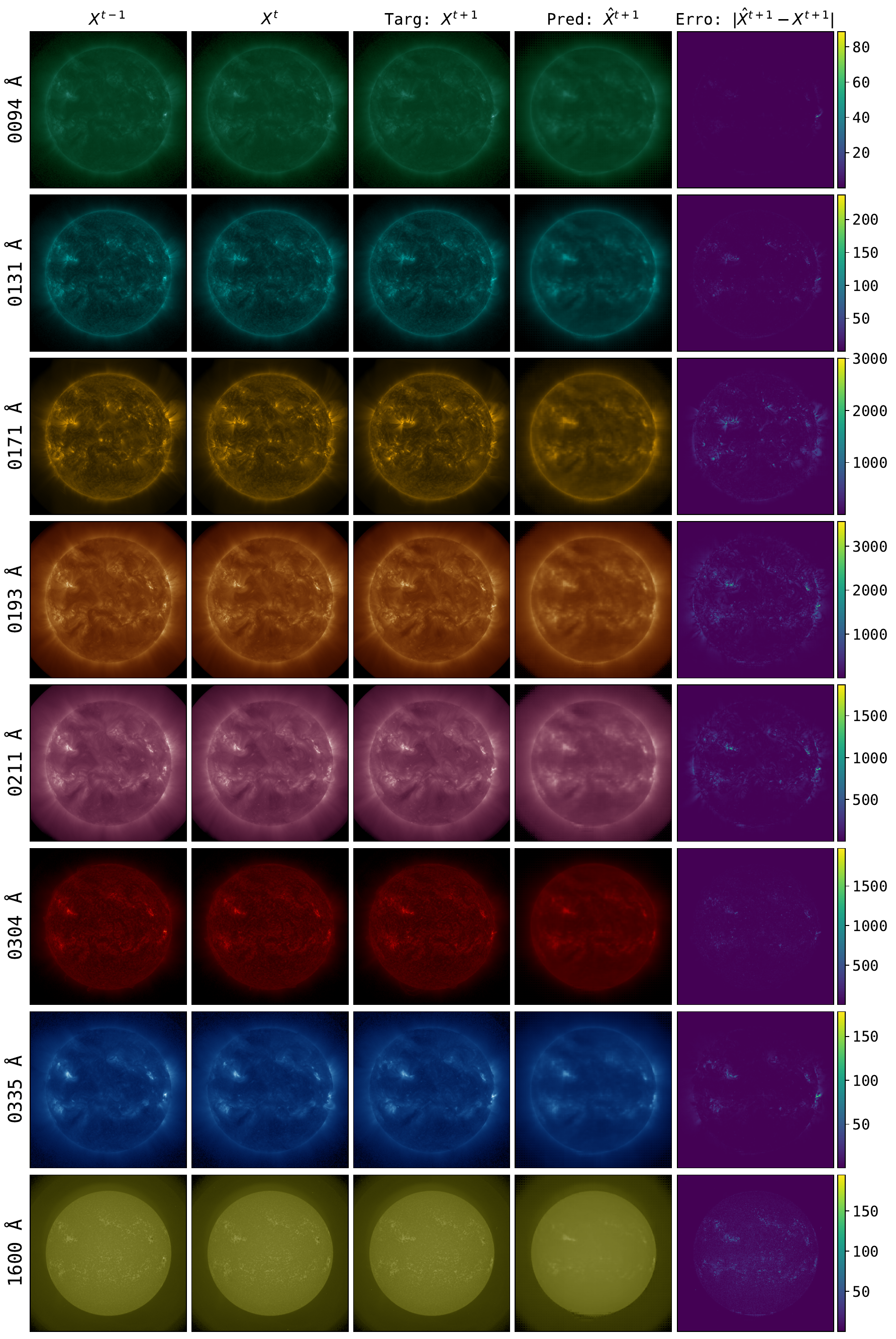}
    \caption{Multi-wavelength forecasting results from $\text{Solaris}_\text{T}$ for 2023-09-02 00:00 UTC. From left to right: input states $X^{t-1}$ and $X^t$, target state $X^{t+1}$, predicted state $\hat{X}^{t+1}$, and absolute error $|\hat{X}^{t+1} - X^{t+1}|$. Rows show results across different wavelengths (94, 131, 171, 193, 211, 304, 335, and 1600 Å).}
    \label{fig:plot_tiny_3}
\end{figure}

\begin{figure}
    \centering
    \includegraphics[width=1.0\linewidth]{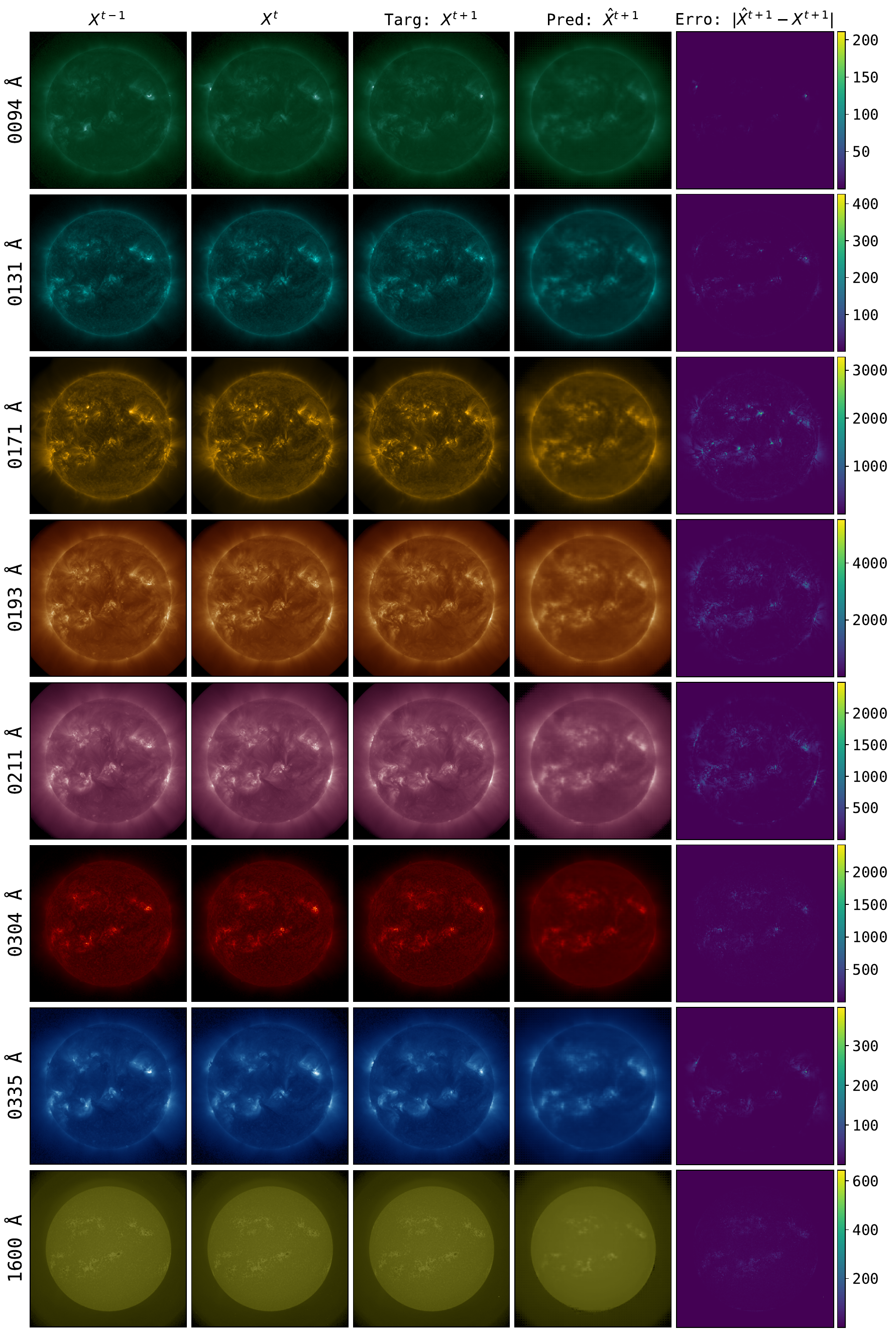}
    \caption{Multi-wavelength forecasting results from $\text{Solaris}_\text{S}$ for 2023-08-02 00:00 UTC. From left to right: input states $X^{t-1}$ and $X^t$, target state $X^{t+1}$, predicted state $\hat{X}^{t+1}$, and absolute error $|\hat{X}^{t+1} - X^{t+1}|$. Rows show results across different wavelengths (94, 131, 171, 193, 211, 304, 335, and 1600 Å).}
    \label{fig:plot_small_2}
\end{figure}

\begin{figure}
    \centering
    \includegraphics[width=1.0\linewidth]{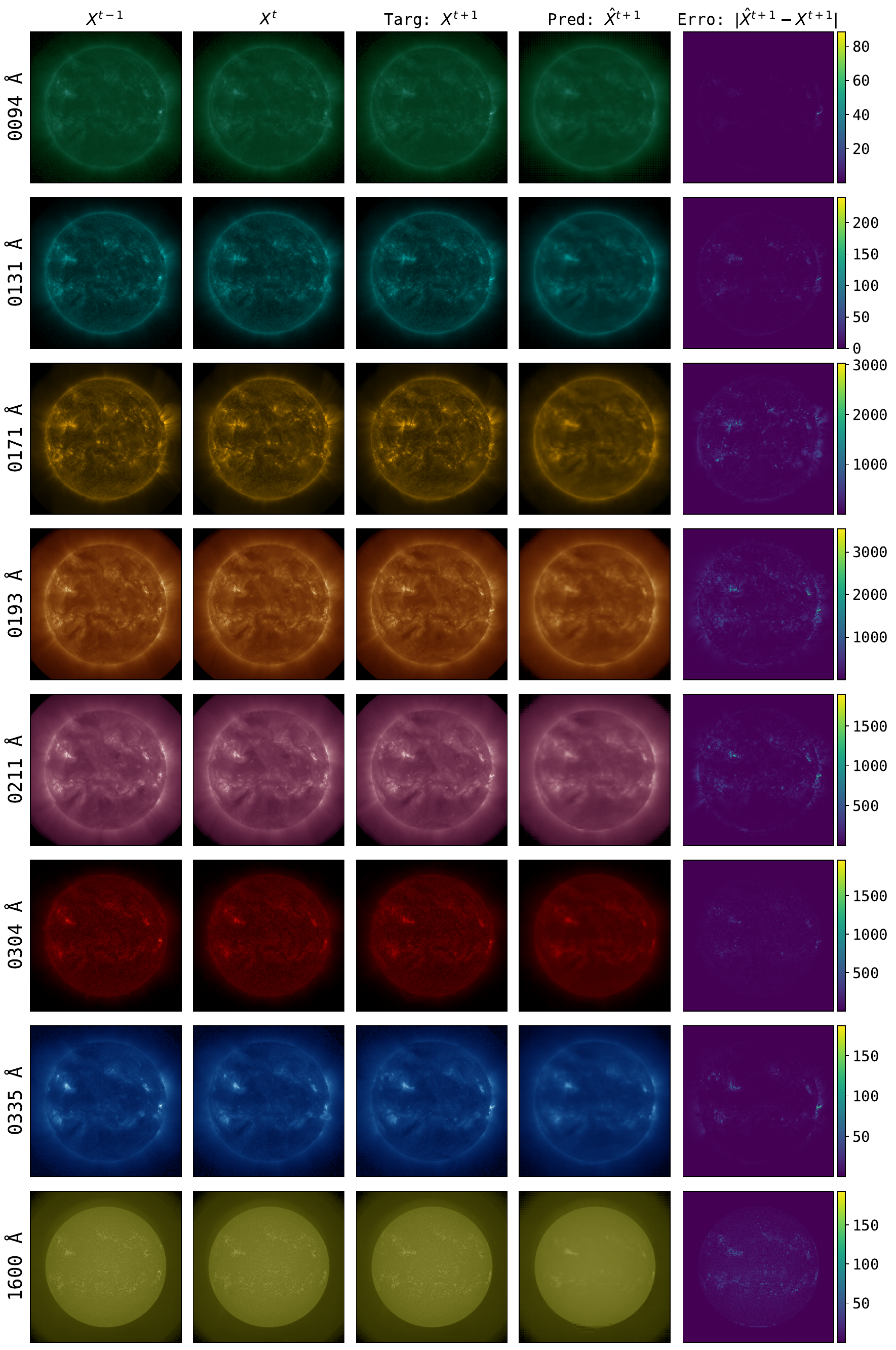}
    \caption{Multi-wavelength forecasting results from $\text{Solaris}_\text{S}$ for 2023-09-02 00:00 UTC. From left to right: input states $X^{t-1}$ and $X^t$, target state $X^{t+1}$, predicted state $\hat{X}^{t+1}$, and absolute error $|\hat{X}^{t+1} - X^{t+1}|$. Rows show results across different wavelengths (94, 131, 171, 193, 211, 304, 335, and 1600 Å).}
    \label{fig:plot_small_3}
\end{figure}

\end{document}